\documentclass[aps,pra,twocolumn,groupedaddress,showpacs,amsfonts,amssymb,superscriptaddress]{revtex4}
\usepackage{graphics}
\usepackage{epsfig}
\usepackage{amsmath}

\begin{document}

% number in the upper righthand corner of the title page in preprint 
% mode.
%\preprint{}

\title{Quantum channels with a finite memory}


\author{Garry Bowen}
\email{gab30@cam.ac.uk}
%\thanks{}
\affiliation{Centre for Quantum Computation, Clarendon Laboratory, 
University of Oxford, Oxford OX1 3PU, United Kingdom.}
\author{Stefano Mancini}
\email{stefano.mancini@unicam.it}
%\thanks{}
\affiliation{Dipartimento di Fisica, Universit\`{a} di Camerino, 
I-62032 Camerino, Italy.}
\date{\today}

\begin{abstract}
In this paper we study 
quantum communication channels with correlated noise effects,
i.e., quantum channels with memory.
We derive a model for correlated noise channels that includes a channel memory state.
We examine the case 
where the memory is finite, and derive bounds on the classical 
and quantum capacities.  For the entanglement-assisted and unassisted classical capacities it is shown that these bounds are attainable for certain classes of channel. Also, we show that the structure of any finite memory state is unimportant in the asymptotic limit, and specifically, for a \textit{perfect} finite-memory channel where no information is lost 
to the environment, achieving the upper bound implies that the channel is asymptotically noiseless.
\end{abstract}

%\keywords{}

\maketitle

\section{Introduction}

Quantum communication through noisy channels has, to date, mainly 
concentrated on quantum channels that are memoryless.  A memoryless 
channel is defined as a noisy channel where the noise acts 
independently on each symbol transmitted through the channel.  In 
classical information theory the discrete memoryless channel (DMC) is 
well understood in terms of the capacity of the channel, and the 
capacity of such channels is invariant under inclusion of feedback or 
shared random correlations \cite{cover}.  The existing capacity 
theorems for quantum channels have also concentrated on the 
memoryless case \cite{lloyd97,schumacher97,holevo98,bennett01a}.  The calculation 
of capacities for classical channels with correlated noise, or 
\textit{memory} channels, has had much more limited success.  One 
type of classical memory channel for which the capacity is known is 
the channel with Markov correlated noise.  In this paper, we examine 
the quantum extension of the channel with Markov correlated noise.  
In particular, we examine a model of a correlated noise 
channel which utilizes unitary operations between the transmitted states, an environment, and a shared memory state, and determine some of the characteristics of such a class of 
quantum channels.

The fact that entangled alphabet states may increase the classical 
capacity of a particular correlated noise channel has been shown 
\cite{macchiavello02}.  For the corresponding memoryless channel the 
capacity is known to be additive, and hence entangled input states cannot 
increase the classical capacity of the memoryless channel over 
product state encoding.

The phrase \textit{finite memory} is used to describe one aspect of the model, a memory state of finite dimension.  The finiteness of the memory is defined only in terms of the model used to 
describe the correlated noise of the channel, and is not necessarily a physical consideration.  The correlations 
between errors may be considered either temporally over each use of a 
single channel, or spatially between uses of many parallel channels.  
The dimension of the memory is determined by the number of Kraus 
operators in the single channel expansion and the correlation length 
of the channel, which may be defined as the maximum number of channel 
uses for which the noise is not conditionally independent.  Any 
channel with a finite correlation length may be generated by a 
channel with a finite memory, according to this model.  Although a 
physical interpretation of the model is not necessary to achieve the 
goal of determining capacity theorems, it may give an understanding 
of the physical motivation.  Over short times the environment with 
which the transmitted state acts may be assumed to be arbitrarily 
large, with interactions between components of the vast environment 
essentially making the recovery of the information impossible.  The 
memory, however, may be interpreted as a subspace of the 
environment which does not ``decay'' over the timescale of separate 
uses of the channel, and is therefore dependent on the previous 
state of the channel.

A physical example of a memory channel is the recent proposal by Bose 
\cite{bose02a}, which uses unmodulated spin chains to transmit 
quantum information.  In this case the initial state introduced to 
the chain by Alice acts as both the input state and a part of the 
memory state for further uses of the channel, as it is assumed Alice 
replaces each transmitted state with a new spin state after each use 
of the channel, whilst the remaining elements of the spin chain 
constitute both the physical channel, the memory of the channel, and 
the output state, which may be removed from the chain for future 
decoding by Bob.

\section{A Model for Quantum Memory Channels}

The Kraus representation theorem \cite{kraus} is an elegant and powerful method of representing quantum dynamics in two different ways, as a sum over operators acting on the state, or alternatively as a unitary evolution of a state and environment.  The unitary interaction model provides an intuitive understanding of open quantum systems, as well as providing a method of calculation.  In deriving our model of a quantum memory channel, we try to preserve the useful aspects that such a unitary representation provides.

\subsection{Unitary Representation of Memoryless Channels}

A quantum channel is defined as a completely positive, trace preserving map from the set of density operators to itself.  Any such map may be represented as a unitary operation between the system state and an environment with a known initial state.  For a single channel use the output state is given by,
\begin{equation}
\rho'_Q = \mathrm{Tr}_E \Big[ U_{QE} \big( \rho_Q \otimes |0_E\rangle \langle 0_E| \big) U_{QE}^{\dag} \Big]
\end{equation}
with $\rho_Q$ the input state, and $\rho'_Q$ the output state.
For a sequence of transmissions through the channel,
\begin{align}
\rho'_Q &= \mathrm{Tr}_{E} \Big[ 
U_{n,E_n}...U_{1,E_1} \big(\rho_Q \otimes |0_{E_1} ... 0_{E_n} \rangle \langle 0_{E_1} ... 0_{E_n} | \big) \nonumber \\
&\phantom{=}\:\times U_{1,E_1}^{\dag} ...  U_{n,E_n}^{\dag}\Big] \\
&= \big(\Lambda_{n} \otimes ... \otimes \Lambda_{1} \big) \rho_Q
\label{eqn:memoryless_model}
\end{align}
where the state $\rho_Q$ now represents a (possibly entangled) input state across the $n$ channel uses, and the environment state is a product state $|0_{E_1} ... 0_{E_n} \rangle = |0_{E_1} \rangle \otimes ... \otimes |0_{E_n} \rangle$.

\subsection{A Unitary Model for Memory Channels}

One model of a quantum memory channel is where each state going 
through the channel acts with a unitary interaction \textit{on the same 
channel memory state}, as well as an independent environment.  The 
backaction of the channel state on the message state therefore gives 
a memory to the channel.
The general model thus includes a channel memory $M$, and the independent 
environments for each qubit $E_i$.  Hence,
\begin{align}
\rho'_Q &= \mathrm{Tr}_{ME} \Big[ 
U_{n,ME_n}...U_{1,ME_1} \big(\rho_Q \otimes |M\rangle \langle M| \nonumber \\
&\phantom{=}\:\otimes |0_{E_1} ... 0_{E_n} \rangle \langle 0_{E_1} ... 0_{E_n} | \big) U_{1,ME_1}^{\dag} ...  U_{n,ME_n}^{\dag}\Big] \nonumber \\
&= \mathrm{Tr}_{M} \Big[ \Lambda_{n,M} ... \Lambda_{1,M} 
\big(\rho_Q \otimes |M\rangle \langle M| \big) \Big]
\label{eqn:memory_model}
\end{align}
where $\rho_Q$ and $\rho'_Q$ are the
input and the output state of the channel while
the trace over the environment is over all environment states.  
If the unitaries factor into independent unitaries acting on the 
memory and the combined state and environment, that is, $U_{n,ME_n} = 
U_{n,E_n}U_M$, then the memory traces out and we have a memoryless 
channel.  If the unitaries reduce to $U_{n,M}$, we can call it a 
\textit{perfect} memory channel, as no information is lost to the 
environment.

The question remains as to what channels can be 
modeled by Eq. (\ref{eqn:memory_model})?  From the Kraus representation theorem \cite{kraus}, we know that for any block of 
length $n$, then any channel acting on the $n$ states may be modeled 
with a memory plus environment of dimension at most $d^{\, 2n}$, for 
$d$ the dimension of the channel.  However, the unitary operation may 
not be factorable into a product of operators acting in the form of 
Eq. (\ref{eqn:memory_model}).

\subsection{Examples of Finite-Memory Channels}

A naive ``memory'' channel can be considered by the two qubit channel 
given by the Kraus operators, $A_0 = \frac{1}{2} (\mathbb{I} \otimes 
\mathbb{I})$, and, $A_1 = \frac{1}{2}(\sigma_Z \otimes \sigma_Z)$, 
and can be modeled by using a memory state that is initially in the 
$|0\rangle\langle 0|$ state, and is the target of a CNOT operation by 
only two qubits before being reset to the initial state.  However, 
this channel is essentially just a memoryless channel in the higher 
dimension space, transmitting qudits of dimension four, and cannot 
therefore be considered useful as a model of a memory channel.  This 
channel also does not fit into the model of Eq. (\ref{eqn:memory_model}) 
as the memory is erased separately after every two qubits.  All such 
channels which may be factored into memoryless channels for some 
finite number of uses may therefore be described using the existing 
properties known for memoryless channels.

A simple example of a perfect memory channel is an extension of the qubit dephasing channel.  For this channel CNOT 
gates operate between the qubits going through the channel and a target memory state,
initially given as $|0_M\rangle \langle 0_M|$, which replaces the environment.  The output states 
have the same reduced density matrices as if they had passed through a memoryless dephasing channel, but the states are also correlated across channel uses, 
that is, a product input state does not necessarily give a product output 
state.  We call this channel the correlated dephasing channel.

Using the unitary SWAP gate to model a channel simply acts as a shift by a single state.  Since 
for this ``shift channel'' the SWAP gate the unitaries act to increment the index for the position of the input states, then on a block of $n$ inputs only the last input state is not recoverable.  Hence the transmission rate for intact states for blocks of size $n$ is simply $1 - 1/n$, which approaches a noiseless channel in the limit $n \rightarrow \infty$.

\subsection{Channels with Markovian Correlated Noise}

An important class of channels that may be represented by the memory channel model are channels with Markovian correlated noise.  A 
Markovian correlated noise channel of length $n$, is of the form,
\begin{align}
\Lambda^{(n)}\rho &= \sum_{i_0,...,i_n} 
p_{i_n|i_{n-1}}p_{i_{n-1}|i_{n-2}}... p_{i_1|i_0} p_{i_0} \nonumber \\
&\phantom{=}\:\times \big( A_{i_n} \otimes ... \otimes A_{i_0}\big) \rho 
\big( A^{\dag}_{i_n} \otimes ... \otimes A^{\dag}_{i_0}\big)
\label{eqn:markov_channel}
\end{align}
where the set $A_{i_k}$ are Kraus operators for single uses of the 
channel on state $k$ \cite{macchiavello02}.
The motivation for looking at Markovian channels is that the 
properties of typical sequences generated from Markovian sources are 
well understood, and the typical sequences of errors generated in Eq. 
(\ref{eqn:markov_channel}) will be directly related to these typical 
sequences.

The correlated dephasing channel may be described using the memory to correlate the dephasing error for 
each qubit, that is, the probability of the $k$th qubit undergoing a 
phase error is determined exactly by whether an error occurred on the previous qubit.  Thus, for 
the correlated dephasing channel with error operators $A_{0_n} = \mathbb{I}^{(n)}$ and $A_{1_n} = \sigma^{(n)}_Z$, acting on the $n$th qubit, the conditional probabilities are given by $p_{k_n|j_{n-1}} = \delta_{jk}$, with an initial probability of error given by $p_0 = p_1 = 1/2$.  This channel may be generated 
using the unitary operation,
\begin{align}
U_{i,M}|\phi^{(i)}\rangle |0_M\rangle &= |\phi^{(i)}\rangle 
|0_M\rangle \label{eqn:cordeph1} \\
U_{i,M}|\phi^{(i)}\rangle |1_M\rangle &= 
\sigma_Z^{(i)}|\phi^{(i)}\rangle |1_M\rangle \label{eqn:cordeph2}
\end{align}
with an initial memory state $|M\rangle = 1/\sqrt{2}(|0\rangle + 
|1\rangle )$.  The equivalence of the controlled phase gate in Eqs. (\ref{eqn:cordeph1}) and (\ref{eqn:cordeph2}) to the use of a CNOT with a memory initially in the $|0_M\rangle\langle 0_M|$ state is obtained by noting $U_{\mathrm{CPHASE}} = (\mathbb{I}\otimes H)U_{\mathrm{CNOT}}(\mathbb{I} \otimes H)$, for $H$ a qubit Hadamard rotation on the memory state.  The channel is asymptotically noiseless, as all 
states with an even number of $|1\rangle$'s are invariant, and 
therefore this subspace may be mapped onto by simply adding a single 
ancilla qubit.  The encoding map from $\mathcal{H}^{\otimes n}$ to 
$\mathcal{H}^{\otimes (n+1)}$, may then transform states with even 
numbers of $|1\rangle$'s to the same state tensored with $|0\rangle$, 
and those states with odd numbers of $|1\rangle$'s to these states 
tensored with $|1\rangle$.  The coded subspace of 
$\mathcal{H}^{\otimes (n+1)}$ is then noiseless.  The rate of 
transmission through $n+1$ uses of the channel is therefore 
$n/(n+1)$, which approaches unity in the limit $n \rightarrow \infty$.

For a general channel with Markovian correlated noise, that is $p_{j|j-1} = 
p_{j|(j-1)(j-2)...i}$ for all $i<j$, the channel may be generated 
using the model given in Eq. (\ref{eqn:memory_model}), where the unitary operator is given by,
\begin{equation}
U_{i,ME_i}|\phi^{(i)}\rangle |j_M\rangle |0_{E_i}\rangle = \sum_k 
\sqrt{p_{k|j}} A_k^{(i)}|\phi^{(i)}\rangle |k_M\rangle 
|j_{E_i}\rangle \; .
\label{eqn:markovmem} 
\end{equation}
The initial memory state determined by the values of the initial 
probability vector for the error operators $[ p_0, p_1,..., p_m ]$, for a family of $m$ operators, by the relationship,
\begin{equation}
\vec{\alpha} = \Gamma^{-1} \vec{p}
\end{equation}
for $\Gamma$ the transition matrix with entries $p_{j|i}$, and 
$\vec{\alpha}$ the squares of the amplitudes for the initial memory 
state $|M\rangle = \sum_j \sqrt{\alpha_j}|j_M\rangle$.  This is, of 
course, provided that the transition matrix is not singular.  For a 
singular matrix we may utilize a different unitary operation $V$ on 
the initial use of the channel, which will not change the asymptotic 
behavior of the channel.  We may also utilize a mixed initial memory 
state $\rho_M = \sum_j \alpha_j |j_M\rangle \langle j_M|$, instead of 
the pure state, without affecting the behavior of the channel.

The derivation of the specific model for the correlated dephasing channel in 
Eqs. (\ref{eqn:cordeph1}) and (\ref{eqn:cordeph2}) differs from the 
prescription given in Eq. (\ref{eqn:markovmem}), in that it does not 
require the extra environment.  The unitary operation on the initial 
states produces orthogonal outputs, whereas in the general case the 
states for each $|j_M\rangle$ in Eq. (\ref{eqn:markovmem}) may not 
necessarily be orthogonal without the environment.  If the output 
state for a given $|j_M\rangle$ in Eq. (\ref{eqn:markovmem}) is orthogonal to all other outputs 
generated by different initial memory states, then the final environment state for this particular output
can ``overlap'' and does not need to be orthogonal to the other 
environment states.  This occurs in the correlated dephasing channel, 
and results in the channel requiring no environment at all.  However, 
it shall be shown that the behavior of these two different channel constructions
is identical, as the actual size of the memory becomes irrelevant in 
the asymptotic limit, provided it is finite.

The noisy channel described by Macchiavello and Palma 
\cite{macchiavello02} may be described in the context of Eq. 
(\ref{eqn:markovmem}), with the error operators given by the identity $A_{0_n} = \mathbb{I}^{(n)}$ and the Pauli spin matrices $A_{1_n} = \sigma^{(n)}_X$, $A_{2_n} = \sigma^{(n)}_Y$, and $A_{3_n} = \sigma^{(n)}_Z$, and the transition matrix elements defined as $p_{k|j} = (1-\mu)p_k + \mu \delta_{jk}$, 
where $\mu$ is a correlation parameter.  The steady state probabilities for this transition matrix are given by the uniform distribution.

\section{Capacities for Finite-Memory Channels}

Having generated a model for the channel, we must address whether 
such a model is instructive in obtaining capacity theorems for the 
channels the model represents.  The existence of a unitary 
representation of an interaction with an environment does allow the 
extension of results from memoryless channels which rely on similar 
arguments, such as the coherent information bound and the quantum 
Fano inequality \cite{schumacher96a}.

\subsection{Results for Classical Capacities}

An upper bound on the classical information that may be sent through 
the memory channel is given by the Holevo bound \cite{kholevo73}.
The maximum mutual information generated between sender and receiver, per channel use, 
for $n$ channels is then given by,
\begin{align}
S^{(n)}_{\max} &= \max_{\{ p_i, \rho^i \}} \frac{1}{n} \bigg[ S\Big( \sum_i p_i 
\mathrm{Tr}_{M} \Big[ \Lambda^{(n)}_M \big(\rho^{i}_Q \otimes \rho_M 
\big) \Big] \Big) \nonumber \\
&\phantom{=}\:- \sum_i p_i S\Big( \mathrm{Tr}_{M} \Big[ \Lambda^{(n)}_M 
\big(\rho^{i}_Q \otimes \rho_M \big) \Big] \Big) \bigg] 
\label{eqn:mem_holevo_upperbound}
\end{align}
where for each $n$, $\Lambda^{(n)}_M = \Lambda_{n,M}...\Lambda_{1,M}$, 
is a channel, and the asymptotic limit is achieved by taking $n 
\rightarrow \infty$.
The ensemble of states $\rho^i_Q = \rho^i_A$ is a set of states generated by the sender, Alice,
for unassisted communication, or $\rho^i_Q = \rho^i_{AB}$ is a set of shared entangled states between sender, Alice, and receiver, Bob, for 
entanglement assisted communication, with the requirement that $\rho_B^i = \rho_B$.  To reduce the number of subscripts, the use of the notation $\rho^i \equiv \rho^i_Q$ for the signal states shall be used for the rest of this section.

The argument for achieving this upper 
bound does not extend easily to the memory channel case.  The problem lies in the 
fact that the coding for the channel cannot be broken up into blocks 
of $n$ uses, as the memory state may be entangled with the previous 
block and thus may not be identical for each block.

The bound in Eq. (\ref{eqn:mem_holevo_upperbound}) is achievable for a class of regular 
Markovian correlated noise channels.  The channels are assumed to be representable by unitary Kraus operators (and are therefore unital), and have initial error probability distributions equal to the steady state probabilities.  The asymptotic use of the channel may be segmented into approximate channels of length $n$.  
That is, by tracing out all other states for each length $n$ segment, 
we obtain a channel where for a total length $l \gg n$ we have $\Lambda^{(l)} 
\approx \Lambda^{(n)} \otimes ... \otimes \Lambda^{(n)}$.  From the 
theory of Markov chains, we know that the approximate channel for 
a product state input is given for a single use by,
\begin{align}
\Lambda^{(1)} \rho &\approx \sum_{i_n} \tilde{p}_{i_{n}} A_{i_n} \rho A_{i_n}^{\dag} \nonumber \\
&= \mathrm{Tr}_M \Lambda^{(1)}_{M} \big( \rho \otimes \tilde{\rho}_M \big)
\label{eqn:single_approx}
\end{align}
where $\tilde{p}_{i_{n}} = \tilde{p}_i$ are the steady state 
probabilities, $\tilde{\rho}_M$ is the memory density matrix with the $\tilde{p}_i$ on the diagonal, and $n$ is taken to be suitably large.  The derivations required for this approximation are shown in the Appendix.  Similarly, with $n$ large, two uses 
the channel are approximated by,
\begin{align}
\Lambda^{(2)} \rho &\approx \sum_{i_n,i_{n-1}} 
p_{i_n|i_{n-1}}\tilde{p}_{i_{n-1}} A_{i_n,i_{n-1}} \rho 
A_{i_n,i_{n-1}}^{\dag} \nonumber \\
&= \mathrm{Tr}_M \Lambda^{(2)}_{M} \big( \rho \otimes \tilde{\rho}_M \big)
\label{eqn:double_approx}
\end{align}
with $A_{i_n,i_{n-1}} = A_{i_n} \otimes A_{i_{n-1}}$, and $\rho$ a possibly entangled input state across the two transmissions through the channel.  This construction may be extended for arbitrary lengths $n$.  In 
the case that the initial distribution $p_{i_0}$ is equal to the 
steady state distribution $p_{i_0} = \tilde{p}_i$, the approximations in Eqs. (\ref{eqn:single_approx}) and (\ref{eqn:double_approx}) become exact.  This is true for all lengths $n$, with $\mathrm{diag}\, \rho_M = \mathrm{diag}\, \tilde{\rho}_M$ always, where $\mathrm{diag}\, \rho$ is the density matrix formed from the diagonal elements of $\rho$.  Therefore \textit{the 
achievable rate is obtained immediately from the Holevo--Schumacher--Westmoreland (HSW) theorem} \cite{schumacher97,holevo98}.

The correlated dephasing channel gives an easy example of the achievablility of the capacity, as for this channel any initial distribution is a steady state probability.  A rate equal to the unassisted classical capacity is achieved using the orthogonal states $\{ |0\rangle, |1\rangle \}$ with a priori probability of $p_i = 1/2$ for this channel, and hence the limit is achieved in this case when $n = 1$.  The entanglement assisted capacity for this channel $C_E = 2$ is, however, only achieved in the asymptotic limit as the block size $n\rightarrow \infty$.

In the case that the initial error probabilities differ from the steady state, much of the derivation above is still applicable.  From the convergence properties of regular 
Markovian sequences, we know that $\mathrm{diag}\, \rho_M \rightarrow 
\tilde{\rho}_M$ as $n$ becomes large, where $\tilde{\rho}_M$ is the 
diagonal density matrix with eigenvalues equivalent to the steady state 
probabilities.  Similarly, for any $\delta > 0$ there exists an $n$ 
for which the total probability of the atypical sequences of Kraus 
operators is less than $\delta$.  This follows from the behavior of 
regular Markovian sources in the Shannon theory \cite{cover}.  The 
contribution to the state $\Lambda^{(n)}\rho$ when the initial probabilities are not the steady state probabilities may therefore be small enough such that the bounds on the 
total probability of error may be made arbitrarily small asymptotically, although at present this remains an open question.

For any channel with a finite memory where the capacity equals the upper bound it may be seen that the exact 
nature of the memory has little effect on the asymptotic behavior of 
the channel.  The correlated dephasing channel, where 
two possible constructions exist each with a different sized memory 
state, is an example.  To analyze the behavior we assume that Bob 
has access to the memory after the block is sent, and as such he can 
measure the information in $M$ as well, then reset the memory to a 
given initial state before the next block.  This gives an achievable 
rate,
\begin{align}
R &\equiv \lim_{n\rightarrow \infty} \max_{\{ p_i, \rho^i \}} 
\frac{1}{n} \bigg[ S\Big( \sum_i p_i \Lambda^{(n)} \big(\rho^{i} 
\otimes \rho_M \big) \Big) \nonumber \\
&\phantom{=}\:- \sum_i p_i S\Big( \Lambda^{(n)} 
\big(\rho^{i} \otimes \rho_M \big) \Big) \bigg] \label{eqn:memrate} \\
&\leq \lim_{n\rightarrow \infty} \max_{\{ p_i, \rho^i \}} 
\frac{1}{n} \bigg[ S\Big( \sum_i p_i \mathrm{Tr}_{M} \Big[ 
\Lambda^{(n)} \big(\rho^{i} \otimes \rho_M \big) \Big] \Big) \nonumber \\
&\phantom{=}\:- \sum_i p_i S\Big( \mathrm{Tr}_{M} \Big[ \Lambda^{(n)} 
\big(\rho^{i} \otimes \rho_M \big) \Big] \Big) + 2 \log_2 d_M 
\bigg] \label{eqn:memory_rate} \\
&= S^{(n)}_{\max} + \frac{2}{n} \log_2 d_{M}
\end{align}
where Eq. (\ref{eqn:memory_rate}) follows from Eq. (\ref{eqn:memrate}) by strong subadditivity and the factor $2 \log_2 d_{M}$ is an upper bound on entropy of 
the memory state living in a space of dimension 
$d_{M}$.  The bound for the rate $R$ of a channel generated 
from tracing both the environment and the memory, is then sandwiched 
by the terms, $nS^{(n)}_{\max} + 2 \log_2 d_M \geq nR \geq 
nS^{(n)}_{\max}$, which would approach the channel capacity for the 
channel including access to the memory, as $n \rightarrow \infty$, 
for any finite memory channel.  The channel capacity is thus only 
affected by the loss of information to the environment, and the loss 
of information into the memory state may be seen to vanish in the 
asymptotic limit.
For a perfect memory channel the channel will be asymptotically noiseless, 
as was shown for the examples of the shift channel and the correlated dephasing 
channel.

\subsection{Results for Quantum Capacities}

The quantum capacities are determined by the maximum asymptotic rates at 
which intact quantum states may be transmitted through a noisy 
quantum channel.  The coherent information bound \cite{schumacher96a,schumacher96} on the quantum capacity applies directly to the case of memory channels.  The role the memory plays in the coherent information bound may be seen by examining the converse to the bound, the quantum Fano inequality, which is shown in the next section.

There exist a number of quantum capacities 
dependent on available additional resources.  Primarily there is the 
unassisted quantum capacity $Q$, the capacities assisted by classical 
side channels $Q_1, Q^{FB}, Q_2$, denoting forward, backward 
(feedback), and two way classical communication respectively, and, 
the entanglement assisted quantum capacity $Q_E$, achievable when 
sender and receiver share unlimited amounts of entanglement prior to 
communication taking place.  For memoryless channels the situation is 
slightly simplified by the equivalence $Q_1 = Q$ 
\cite{bennett96,barnum00}, whether this holds for channels with 
memory is not yet known.

The entanglement assisted quantum capacity is simply related to the 
entanglement assisted classical capacity by the use of quantum dense 
coding and quantum teleportation, giving the equality $C_E = 2Q_E$ 
\cite{bennett99}.  The actual nature of the channel does not affect this relationship.

\subsection{The Quantum Fano Inequality}

The quantum Fano inequality \cite{schumacher96a} is used to give a 
converse to any quantum coding theorems.  The inequality describes 
the loss in fidelity of the transmitted states that occurs due to the 
exchange of entropy to the environment during transmission through 
the channel.

Taking a state $\rho_Q$ with a purification in terms of a reference system $R$, such that, $\rho_Q = \mathrm{Tr}_R |\psi_{QR}\rangle \langle \psi_{QR}|$, we define the entanglement fidelity as $F = \langle \psi_{QR}|\rho'_{QR}|\psi_{QR}\rangle$, where $\rho'_{QR}$ is the total output state following the transmission of $\rho_Q$ through the noisy channel.
The quantum Fano inequality may be applied to the finite memory channel by 
simply noting that the entropy exchange to the environment $E$ may be rewritten as,
\begin{equation}
S(\rho'_{E}) = S(\rho'_{MQR}) \leq S(\rho'_{M})+S(\rho'_{QR}) \; .
\end{equation}
It is assumed here that the memory state is initially pure, as it does not affect the derivation compared to a mixed memory state.  This is because any finite memory state may be purified with another finite reference system.  This is also equivalent to applying the Fano inequality using $S(\rho'_{ME})$ as the environment, and then utilizing the Araki--Lieb inequality to obtain $S(\rho'_{QR}) \geq S(\rho'_{E}) - S(\rho'_{M})$.

This leads to a Fano 
inequality for channels with a finite memory,
\begin{equation}
S_E \leq \log_{2} d_M + h(F) + (1-F)\log_{2} (d^2 - 1)
\end{equation}
for $S_E$ the entropy exchange with the environment, $F$ the entanglement fidelity, 
$h(F) = - F \log_2 F - (1-F) \log_2 (1-F)$ the binary entropy of the entanglement fidelity, $d$ the 
dimension of $\mathcal{H}_Q$, and $d_M$ the dimension of the memory.  
For a single channel use, this inequality may be weak, but in the 
case of multiple uses the inequality can become stronger.  This is 
due to the average entropy exchange for a large number of channel uses $N$ being given by,
\begin{align}
\frac{1}{N}S_E &\leq \frac{1}{N}\left[ \log_{2} d_M + h(F) 
+ (1-F)\log_{2} (d^{\, 2N}-1) \right] \nonumber \\
&\approx 2(1-F)\log_{2} d
\end{align}
where the first two terms in the sum on the right hand side may be 
made arbitrarily small, given large enough $N$.  This may be 
interpreted as the fact that a high entanglement fidelity over many 
uses of the channel necessarily implies a low average entropy 
exchange with the environment.  In the asymptotic limit the particular channel construction used, and the exact nature of any finite memory state, are both ``irrelevant'' in terms of the bounds on the channel capacity.

\section{Conclusion}

A model for a class of quantum channels with memory has been 
proposed.  The class of channels that may be described by this model 
include the set of channels with Markovian correlated quantum noise.  
For these types of channels it has 
been shown that the memory state required to generate the channel is 
finite.  These finite memory channels have similar asymptotic 
behavior to the quantum memoryless channels, in that they may be essentially 
described by the loss of information to an initial product state 
environment after a unitary interaction with the states transmitted 
through the channel.  The size of the memory state is finite, and so 
the effect on loss from the channel is vanishing in the asymptotic 
limit.  The simplest demonstration is the case of perfect memory 
channels where no information at all is lost to the environment and 
so achievement of the upper bound on the capacity for this class of channels will
asymptotically give a noiseless 
quantum channel.

It has also been demonstrated that Holevo--Schumacher--Westmoreland coding
can achieve the capacity bound for channels with Markov correlated noise, where
the Kraus operators are unitary, providing the initial error probabilities are
equal to the steady state probabilities for the regular Markov chain.

The unitary representation of the channel also 
allows for derivations of bounds on the quantum capacity using the 
coherent information, and application of the quantum Fano inequality to 
finite memory channels.

\appendix*
\section{Evolution of the channel and memory state}
\subsection{Derivation of the Channel from the Unitary Construction}

Here it is shown that the diagonal elements of the memory state determine the error operators for the next transmitted state.  For the memory state $\rho_M = \sum_{jl} \lambda_{jl} \, |j_M\rangle \langle l_M|$, the channel for the next transmitted state is given by,
\begin{equation}
\begin{split}
&\sum_{jl} \lambda_{jl} \mathrm{Tr}_{ME} \big[ U_{QME}|\phi_Q\rangle |j_M\rangle |0_{E}\rangle \langle 0_{E}| \langle l_M| \langle \phi_Q| U_{QME}^{\dag} \big] \\
&= \sum_{jl} \lambda_{jl} \mathrm{Tr}_M \bigg[ \sum_{km} 
\sqrt{p_{k|j}p_{m|l}}\delta_{jl} A_k|\phi_Q\rangle |k_M\rangle \langle m_M| \langle \phi_Q|A_{m}^{\dag} \bigg] \nonumber \\
&= \sum_{jl} \lambda_{jl}\delta_{jl} \sum_{kmn} 
\sqrt{p_{k|j}p_{m|l}} A_k|\phi\rangle \langle n|k_M\rangle \langle m_M|n\rangle \langle \phi|A_{m}^{\dag} \nonumber \\
&= \sum_{j} \lambda_{jj} \sum_{k} p_{k|j} A_k |\phi_Q\rangle \langle \phi_Q|A_{k}^{\dag} \nonumber \\
\end{split}
\end{equation}
The error operators are determined by the diagonal elements of the memory only, the off-diagonal matrix elements have no effect on the channel.

\subsection{Convergence of the Diagonal Elements of the Memory State}

The exact nature of the memory state itself depends on the states transmitted through the channel.  Perhaps surprisingly, however, the diagonal elements of the state are \textit{independent} of the transmitted states.  To show this we note for a memory initially in the state $\rho_M = \sum_{jl} \lambda_{jl} \, |j_M\rangle \langle l_M|$, the new memory state after one iteration of the channel is,
%\begin{widetext}
\begin{equation}
\begin{split}
&\sum_{jl} \lambda_{jl} \mathrm{Tr}_{QE} \big[ U_{QME}|\phi_Q\rangle |j_M\rangle |0_{E}\rangle \langle 0_{E}| \langle l_M| \langle \phi_Q| U_{QME}^{\dag} \big] \\
&= \sum_{jl} \lambda_{jl} \mathrm{Tr}_Q \bigg[ \sum_{kmn} 
\sqrt{p_{k|j}p_{m|l}}\delta_{jl} A_k|\phi_Q\rangle |k_M\rangle \langle m_M| \langle \phi_Q|A_{m}^{\dag} \bigg] \nonumber \\
&= \sum_{jl} \lambda_{jl}\delta_{jl} \sum_{kmn} 
\sqrt{p_{k|j}p_{m|l}} \langle n|A_k|\phi\rangle |k_M\rangle \langle m_M| \langle \phi|A_{m}^{\dag}|n\rangle \nonumber \\
&= \sum_{j} \lambda_{jj} \sum_{km} 
\sqrt{p_{k|j}p_{m|j}} \langle \phi|A_{m}^{\dag}A_k|\phi \rangle |k_M\rangle \langle m_M| \nonumber \\
&= \sum_{j} \lambda_{jj} \bigg[ \sum_{k} 
p_{k|j} \langle \phi|A_{k}^{\dag}A_k|\phi \rangle |k_M\rangle \langle k_M| \nonumber \\
&\phantom{=}\: + \sum_{k\neq m} \sqrt{p_{k|j}p_{m|j}} \langle \phi|A_{m}^{\dag}A_k|\phi \rangle |k_M\rangle \langle m_M| \bigg]
\end{split}
\end{equation}
%\end{widetext}
if $m = k$ then the unitaries give $A_{m}^{\dag}A_k = \mathbb{I}$, therefore the diagonal elements of $\rho_M$ undergo the process of a Markov chain.  The off-diagonal elements do not necessarily vanish, but they do not affect the error operators acting on the transmitted states.  Only the diagonal elements of the memory state affect the behavior of the channel.
\begin{acknowledgments}
While this work was undertaken, GB was supported by the Oxford-Australia Trust, the Harmsworth Trust, 
and Universities UK.
\end{acknowledgments}

%\bibliography{references.bib}

\begin{thebibliography}{14}
\expandafter\ifx\csname natexlab\endcsname\relax\def\natexlab#1{#1}\fi
\expandafter\ifx\csname bibnamefont\endcsname\relax
  \def\bibnamefont#1{#1}\fi
\expandafter\ifx\csname bibfnamefont\endcsname\relax
  \def\bibfnamefont#1{#1}\fi
\expandafter\ifx\csname citenamefont\endcsname\relax
  \def\citenamefont#1{#1}\fi
\expandafter\ifx\csname url\endcsname\relax
  \def\url#1{\texttt{#1}}\fi
\expandafter\ifx\csname urlprefix\endcsname\relax\def\urlprefix{URL }\fi
\providecommand{\bibinfo}[2]{#2}
\providecommand{\eprint}[2][]{\url{#2}}

\bibitem[{\citenamefont{Cover and Thomas}(1991)}]{cover}
\bibinfo{author}{\bibfnamefont{T.~M.} \bibnamefont{Cover}} \bibnamefont{and}
  \bibinfo{author}{\bibfnamefont{J.~A.} \bibnamefont{Thomas}},
  \emph{\bibinfo{title}{Elements of Information Theory}}
  (\bibinfo{publisher}{Wiley}, \bibinfo{address}{New York},
  \bibinfo{year}{1991}).

\bibitem[{\citenamefont{Lloyd}(1997)}]{lloyd97}
\bibinfo{author}{\bibfnamefont{S.}~\bibnamefont{Lloyd}},
  \bibinfo{journal}{Phys. Rev. A} \textbf{\bibinfo{volume}{55}},
  \bibinfo{pages}{1613} (\bibinfo{year}{1997}).

\bibitem[{\citenamefont{Schumacher and Westmoreland}(1997)}]{schumacher97}
\bibinfo{author}{\bibfnamefont{B.}~\bibnamefont{Schumacher}} \bibnamefont{and}
  \bibinfo{author}{\bibfnamefont{M.~D.} \bibnamefont{Westmoreland}},
  \bibinfo{journal}{Phys. Rev. A} \textbf{\bibinfo{volume}{56}},
  \bibinfo{pages}{131} (\bibinfo{year}{1997}).

\bibitem[{\citenamefont{Holevo}(1998)}]{holevo98}
\bibinfo{author}{\bibfnamefont{A.~S.} \bibnamefont{Holevo}},
  \bibinfo{journal}{IEEE Trans. Inform. Theory} \textbf{\bibinfo{volume}{44}},
  \bibinfo{pages}{269} (\bibinfo{year}{1998}).

\bibitem[{\citenamefont{Bennett et~al.}(2002)\citenamefont{Bennett, Shor,
  Smolin, and Thapliyal}}]{bennett01a}
\bibinfo{author}{\bibfnamefont{C.~H.} \bibnamefont{Bennett}},
  \bibinfo{author}{\bibfnamefont{P.~W.} \bibnamefont{Shor}},
  \bibinfo{author}{\bibfnamefont{J.~A.} \bibnamefont{Smolin}},
  \bibnamefont{and} \bibinfo{author}{\bibfnamefont{A.~V.}
  \bibnamefont{Thapliyal}}, \bibinfo{journal}{IEEE Trans. Inform. Theory}
  \textbf{\bibinfo{volume}{48}}, \bibinfo{pages}{2637} (\bibinfo{year}{2002}),
  \eprint{quant-ph/0106052}.

\bibitem[{\citenamefont{Macchiavello and Palma}(2002)}]{macchiavello02}
\bibinfo{author}{\bibfnamefont{C.}~\bibnamefont{Macchiavello}}
  \bibnamefont{and} \bibinfo{author}{\bibfnamefont{G.~M.} \bibnamefont{Palma}},
  \bibinfo{journal}{Phys. Rev. A} \textbf{\bibinfo{volume}{65}},
  \bibinfo{pages}{050301R} (\bibinfo{year}{2002}), \eprint{quant-ph/0107052}.

\bibitem[{\citenamefont{Bose}()}]{bose02a}
\bibinfo{author}{\bibfnamefont{S.}~\bibnamefont{Bose}},
\bibinfo{journal}{Phys. Rev. Lett.} \textbf{\bibinfo{volume}{91}},
  \bibinfo{pages}{207901} (\bibinfo{year}{2003}), \eprint{quant-ph/0212041}.

\bibitem[{\citenamefont{Kraus}(1983)}]{kraus}
\bibinfo{author}{\bibfnamefont{K.}~\bibnamefont{Kraus}},
  \emph{\bibinfo{title}{States, Effects, and Operations: Fundamental Notions of
  Quantum Theory}} (\bibinfo{publisher}{Springer--Verlag},
  \bibinfo{address}{Berlin}, \bibinfo{year}{1983}).

\bibitem[{\citenamefont{Schumacher}(1996)}]{schumacher96a}
\bibinfo{author}{\bibfnamefont{B.}~\bibnamefont{Schumacher}},
  \bibinfo{journal}{Phys. Rev. A} \textbf{\bibinfo{volume}{54}},
  \bibinfo{pages}{2614} (\bibinfo{year}{1996}).

\bibitem[{\citenamefont{Holevo}(1973)}]{kholevo73}
\bibinfo{author}{\bibfnamefont{A.~S.} \bibnamefont{Holevo}},
  \bibinfo{journal}{Probl. Peredachi Inf.} \textbf{\bibinfo{volume}{9}},
  \bibinfo{pages}{3} (\bibinfo{year}{1973}).

\bibitem[{\citenamefont{Schumacher and Nielsen}(1996)}]{schumacher96}
\bibinfo{author}{\bibfnamefont{B.}~\bibnamefont{Schumacher}} \bibnamefont{and}
  \bibinfo{author}{\bibfnamefont{M.~A.} \bibnamefont{Nielsen}},
  \bibinfo{journal}{Phys. Rev. A} \textbf{\bibinfo{volume}{54}},
  \bibinfo{pages}{2629} (\bibinfo{year}{1996}).

\bibitem[{\citenamefont{Bennett et~al.}(1996)\citenamefont{Bennett, DiVincenzo,
  Smolin, and Wootters}}]{bennett96}
\bibinfo{author}{\bibfnamefont{C.~H.} \bibnamefont{Bennett}},
  \bibinfo{author}{\bibfnamefont{D.~P.} \bibnamefont{DiVincenzo}},
  \bibinfo{author}{\bibfnamefont{J.~A.} \bibnamefont{Smolin}},
  \bibnamefont{and} \bibinfo{author}{\bibfnamefont{W.~K.}
  \bibnamefont{Wootters}}, \bibinfo{journal}{Phys. Rev. A}
  \textbf{\bibinfo{volume}{54}}, \bibinfo{pages}{3824} (\bibinfo{year}{1996}).

\bibitem[{\citenamefont{Barnum et~al.}(2000)\citenamefont{Barnum, Knill, and
  Nielsen}}]{barnum00}
\bibinfo{author}{\bibfnamefont{H.}~\bibnamefont{Barnum}},
  \bibinfo{author}{\bibfnamefont{E.}~\bibnamefont{Knill}}, \bibnamefont{and}
  \bibinfo{author}{\bibfnamefont{M.~A.} \bibnamefont{Nielsen}},
  \bibinfo{journal}{IEEE Trans. Inform. Theory} \textbf{\bibinfo{volume}{46}},
  \bibinfo{pages}{1317} (\bibinfo{year}{2000}).

\bibitem[{\citenamefont{Bennett et~al.}(1999)\citenamefont{Bennett, Shor,
  Smolin, and Thapliyal}}]{bennett99}
\bibinfo{author}{\bibfnamefont{C.~H.} \bibnamefont{Bennett}},
  \bibinfo{author}{\bibfnamefont{P.~W.} \bibnamefont{Shor}},
  \bibinfo{author}{\bibfnamefont{J.~A.} \bibnamefont{Smolin}},
  \bibnamefont{and} \bibinfo{author}{\bibfnamefont{A.~V.}
  \bibnamefont{Thapliyal}}, \bibinfo{journal}{Phys. Rev. Lett.}
  \textbf{\bibinfo{volume}{83}}, \bibinfo{pages}{3081} (\bibinfo{year}{1999}).

\end{thebibliography}

\end{document}